\documentclass[12pt]{article}
\usepackage[dvips]{graphicx}
\usepackage{amsmath,amssymb,amsthm}

\def\Bx{{\bf x}}
\def\By{{\bf y}}
\def\Ba{{\bf a}}
\def\Bb{{\bf b}}

\def\Bbeta{{\bf \beta}}

\newtheorem{proposition}{Proposition}[section]
\newtheorem{example}{Example}[section]
\newtheorem{theorem}{Theorem}[section]
\newtheorem{definition}{Definition}[section]
\newtheorem{conjecture}{Conjecture}[section]

\title{Design and analysis of fractional factorial experiments from the viewpoint of computational 
algebraic statistics}
\date{April, 2010}
\author{Satoshi Aoki%
\thanks{Graduate School of Science and Engineering (Science Course), Kagoshima University.}%
\ \thanks{JST, CREST.}
and Akimichi Takemura%
\thanks{Department of 
Mathematical Informatics, 
Graduate School of Information Science and Technology, 
University of Tokyo.}\ \footnotemark[2]
}

\begin{document}
\maketitle

\begin{abstract}
We give an expository review of
applications of computational algebraic statistics to design and
analysis of fractional factorial experiments  
based on our recent works.
For the purpose of design,
the techniques of Gr\"obner bases and indicator functions allow us to treat
fractional factorial designs without distinction between regular designs and 
non-regular designs.
For the purpose of analysis of data from fractional factorial designs, the
techniques of Markov bases allow us to handle discrete observations.
Thus the approach of computational algebraic statistics greatly enlarges
the scope  of fractional factorial designs.
\end{abstract}

\section{Introduction}
\label{sec:intro}
Application of Gr\"obner bases theory to  designed experiments is
an attractive topic in a relatively new field in statistics,
called {\it computational algebraic statistics}. After the first work by
Pistone and Wynn (\cite{Pistone-Wynn-1996}) this topic is vigorously
studied both by algebraists and statisticians.
These  developments are  stimulated by advancements
in algebraic algorithms. By recent algorithms some practical computations 
are becoming feasible for statistical applications. For these backgrounds see
\cite{Aoki-Takemura-2009-trans}. 

In this paper we revisit some fundamental results in this field, mainly
in the treatments of two-level fractional factorial designs.
The most important part of the long history of 
studies of two-level fractional factorial designs
is the theory of regular designs. 
As explained in many classical works (e.g.\ 
\cite{Box-Hunter-1961},  \cite{Box-Hunter-Hunter-1978}), 
properly chosen regular fractional factorial designs have many desirable
properties, mainly due to orthogonality and balancedness of  designs. 
In addition, an elegant theory based on the linear algebra over 
$GF(2)$ is well established for regular two-level fractional factorial
designs (e.g.\ \cite{Mukerjee-Wu-2006}). 

On the other hand, there are still 
many open problems 
on general structures of non-regular fractional
factorial designs, except for some specific designs 
such as Plackett-Burman designs.
In computational algebraic statistics, designs are simply 
characterized as the solutions of a set of polynomial
equations. Therefore we need not distinguish between regular designs and
non-regular designs.
In fact, this algebraic treatment yields many new results. 
For example,  concepts defined for regular designs,
such as resolution and aberration,
can be naturally generalized to
non-regular designs in the framework of computational algebraic statistics.

The construction of this paper is as follows.
In Section \ref{sec:design-ideal} we review some algebraic definitions for
handling  fractional factorial designs. In Section
\ref{sec:confounding-ideal-membership} we present a method of representing the
confounding relations between factor effects algebraically,
which is one of the results in \cite{Pistone-Wynn-1996}.
In Section \ref{sec:indicator-function} an indicator function 
defined in \cite{Fontana-Pistone-Rogantin-2000} is given. The indicator function
is a valuable tool to characterize non-regular designs. 
In Section \ref{sec:adding-factor} we show a simple result related  to the
indicator function when new factors are added or when interaction effects 
are formally considered as factors.
In Section \ref{sec:opt-design-selection} we
consider the problem of selecting optimal designs.
In Section \ref{sec:markov-basis} 
we discuss Markov chain Monte Carlo approach
for testing factor effects, when observations are discrete random variables.
We end the paper with some discussions in Section \ref{sec:discussions}.

Throughout this paper, we use terminology of Gr\"obner bases theory without definition. In
particular, we omit explanations for term  orders and division
algorithms. For these notions, see 
\cite{Cox-Little-O'Shea-1997} or \cite{Adams-Loustaunau-1994}, for
example.

\section{Design ideals}
\label{sec:design-ideal}
Consider fractional factorial designs of $m$ controllable factors.
We assume that the levels of each factor are coded as elements of a field $K$, which is
a finite extension of the field 
$\mathbb{Q}$ of rational numbers. 
For the case of factors with two levels we can take $K=\mathbb{Q}$.
However for the case of factors with more than two levels,
it is advantageous to code the levels by complex numbers (\cite{Pistone-Rogantin-2008a},
\cite{Pistone-Rogantin-2008b}).  In this case $K$ contains some complex roots of unity.

A fractional factorial design (without replication) is identified with a finite subset of
$K^m$. In  computational algebraic statistics, this set is
considered as the set of solutions of polynomial equations, 
called an {\it algebraic variety}, and 
the set of polynomials vanishing on
all the solutions is called an {\it ideal}. 
For the rest of this section we only consider the
case of two-level factors. For more general case see 
\cite{Pistone-Riccomagno-Wynn-2001}.

The full factorial design of $m$ factors with two levels is expressed as
\[
{\cal D} = \{(x_1,\ldots,x_m)\ |\ x_1^2 = \cdots =
x_m^2 = 1\} = \{-1,+1\}^m,
\]
where we write $-1$ and $1$ as the two levels. We call a subset 
${\cal F} \subset {\cal D}$ a fractional factorial design.
Let $K[x_1,\ldots,x_m]$ be the polynomial ring of indeterminates
$x_1,\ldots,x_m$ with the coefficients in $K$. Then the set of
polynomials vanishing on the points of ${\cal F}$
\[
 I({\cal F}) = \{f \in K[x_1,\ldots,x_m]\ |\ f(x_1,\ldots,x_m) = 0\
 \mbox{for all}\ (x_1,\ldots,x_m)\in {\cal F}\}
\]
is the design ideal of ${\cal F}$. 

An ideal $I \subset K[x_1,\ldots,x_m]$ is generated by a 
(finite) basis $\{g_1,\ldots,g_k\} \subset  I$ if 
for any $f \in I$ there exist polynomials 
$s_1,\ldots,s_k \in K[x_1,\ldots,x_m]$ such that 
\[
 f(x_1,\ldots,x_m) = \sum_{i = 1}^k s_i(x_1,\ldots,x_m)g_i(x_1,\ldots,x_m).
\]
The above $s_1,\ldots,s_k$ are not unique in general. We write 
$I = \langle g_1,\ldots,g_k \rangle$ if $I$ is generated by a basis
$\{g_1,\ldots,g_k\}$. For example,  for the full factorial design of two
factors with two levels ($2^2$-design), the design ideal of 
${\cal D} = \{-1,+1\}^2$ is written as
\[
 I({\cal D}) = \langle x_1^2-1,x_2^2-1 \rangle .
\]
Every ideal has a finite basis by the Hilbert basis
theorem. In addition, if 
$\{g_1,\ldots,g_k\}$ is a basis of $I({\cal F})$, then ${\cal F}$ coincides
with the solutions of the polynomial equations
$g_1 = 0,\ldots, g_k = 0$.

Suppose there are $n$ runs (i.e.\  points) in a fractional factorial
design ${\cal F} \subset {\cal D}$. A general method to derive a basis
of $I({\cal F})$ is to make use of the algorithm for calculating the
intersection of the ideals. By definition, the design ideal of the 
design consisting of a single point, $(a_1,\ldots,a_m) \in \{-1,+1\}^m$, is written 
as 
\[
 \langle x_1-a_1,\ldots,x_m-a_m \rangle \subset K[x_1,\ldots,x_m].
\]
Therefore the design ideal of the $n$-runs design, 
${\cal F} = \{(a_{i1},\ldots,a_{im}),\ i = 1,\ldots,n\}$, is given as
\begin{equation}
 I({\cal F}) = \bigcap_{i = 1}^n \langle x_1-a_{i1},\ldots,x_m-a_{im}\rangle.
\label{eqn:intersection}
\end{equation}
To calculate the intersection of ideals, we can use the theory of
Gr\"obner bases. In fact, by introducing the indeterminates
$t_1,\ldots,t_n$ and the polynomial ring 
$K[x_1,\ldots,x_m,t_1,\ldots,t_n]$, equation
(\ref{eqn:intersection}) is written as
\begin{equation*}
 I({\cal F}) = I^{*} \cap K[x_1,\ldots,x_m],
\end{equation*}
where
\begin{equation}
 I^{*} = \langle t_i(x_1-a_{i1}),\ldots,t_i(x_m-a_{im}), \; i =
 1,\ldots,n, \ t_1+\cdots + t_n - 1 \rangle
\label{eqn:I^*}
\end{equation}
is an ideal of $K[x_1,\ldots,x_m,t_1,\ldots,t_n]$.
Therefore we can obtain a basis of $I({\cal F})$ as 
the reduced Gr\"obner basis of $I^{*}$
with respect to a 
term order satisfying $\{t_1,\ldots,t_n\}\succ \{x_1,\ldots,x_m\}$.
This argument is known as the elimination theory, one of the important
applications of  Gr\"obner bases (\cite{Cox-Little-O'Shea-1997}).

\begin{example}[A $2^{7-4}_{{\rm III}}$ design] 
\label{exam:2^7-4}
Consider the design known as the orthogonal array $L_8(2^7)$ of 
resolution III with the defining relations 
\begin{equation}
x_3 = -x_1x_2,\ x_5=-x_1x_4,\ x_6=-x_2x_4,\ x_7=x_1x_2x_4
\label{eqn:def-relation}
\end{equation}
given as follows.
\begin{center}
{\rm
\begin{tabular}{crrrrrrr}
run$\backslash$factor & $x_1$ & $x_2$ & $x_3$ & $x_4$ & $x_5$ & $x_6$ &
 $x_7$\\ \hline
1 &-1 &-1 & -1 &-1 &-1 &-1 &-1\\
2 &-1 &-1 & -1 & 1 & 1 & 1 & 1\\ 
3 &-1 & 1 &  1 &-1 &-1 & 1 & 1\\
4 &-1 & 1 &  1 & 1 & 1 &-1 &-1\\
5 & 1 &-1 &  1 &-1 & 1 &-1 & 1\\
6 & 1 &-1 &  1 & 1 &-1 & 1 &-1\\
7 & 1 & 1 & -1 &-1 & 1 & 1 &-1\\
8 & 1 & 1 & -1 & 1 &-1 &-1 & 1\\ \hline
\end{tabular}
}
\end{center}
For this design a basis of $I({\cal F})$ is obtained 
by omitting elements containing the indeterminates
$t_1,\ldots,t_8$ from the reduced Gr\"obner basis of (\ref{eqn:I^*}).
For the lexicographic term order with $x_1\succ\cdots\succ x_7$, the reduced
 Gr\"obner basis is given as 
\begin{equation}
\{x_7^2-1, x_6^2-1, x_5^2-1, x_3+x_5x_6, x_2+x_5x_7,
 x_1+x_6x_7, x_4-x_5x_6x_7 
\},
\label{eqn:gb-lex}
\end{equation}
while for the graded reverse lexicographic term order, the reduced
 Gr\"obner basis is given as   
\begin{equation}
\begin{array}{l}
\{x_7^2-1, x_6^2-1, x_5^2-1, x_4^2-1, x_3^2-1,
 x_2^2-1, x_1^2-1,\\
 x_2x_3+x_1, x_4x_5+x_1, x_6x_7+x_1,
x_1x_3+x_2, x_4x_6+x_2, x_5x_7+x_2,\\
 x_1x_2+x_3, x_4x_7+x_3, x_5x_6+x_3,
x_1x_5+x_4, x_2x_6+x_4, x_3x_7+x_4,\\
 x_1x_4+x_5, x_2x_7+x_5, x_3x_6+x_5,
x_1x_7+x_6, x_2x_4+x_6, x_3x_5+x_6,\\
 x_1x_6+x_7, x_2x_5+x_7, x_3x_4+x_7\}.
\end{array}
\label{eqn:gb-grevlex}
\end{equation}
\end{example}

Hereafter, we write a  monomial of the indeterminates $x_1,\ldots,x_m$ as 
$\Bx^{\Ba} = x_1^{a_1}\cdots x_m^{a_m}$. It is sufficient to consider 
$\Ba = (a_1,\ldots,a_m) \in \{0,1\}^m$ since the two levels are coded
as $\{-1,+1\}$. The results in Example \ref{exam:2^7-4} indicate the
relation between the defining relation and the design ideal for 
regular fractional factorial designs, i.e., designs obtained from the
defining relations such as (\ref{eqn:def-relation}). In fact, for the
design ${\cal F}$ obtained by the defining relation 
\[
 \Bx^{\Ba_\ell} = c_{\ell},\ c_{\ell} \in \{-1,1\}, \ \ell = 1,\ldots,s,
\]
the design ideal is written as
\[
 I({\cal F}) =
 \langle x_1^2-1,\ldots,x_m^2-1,\Bx^{\Ba_1}-c_1,\ldots,\Bx^{\Ba_s}-c_s
\rangle .
\]
For example, the design ideal in Example \ref{exam:2^7-4} is also
written as 
\begin{equation}
 I({\cal F}) =
 \langle x_1^2-1,\ldots,x_7^2-1,x_1x_2x_3+1,x_1x_4x_5+1,x_2x_4x_6+1,x_1x_2x_4x_7-1\rangle .
\label{eqn:basis-2^7-4}
\end{equation}
As we see here, an obvious basis of the design ideal of a
regular fractional factorial design ${\cal F} \subset {\cal D}$ 
consists of defining relations in addition to $x_1^2-1,\ldots,x_m^2-1$. 
Also for non-regular designs we can consider the set
of polynomials (in addition to $x_1^2-1,\ldots,x_m^2-1$) 
which forms a basis of $I({\cal F})$.
This set of the polynomials is called a set of defining equations of
${\cal F}$ in \cite{Fontana-Pistone-Rogantin-2000}. This is a
generalized concept of defining relations from  regular to
non-regular designs.
Note that the above {\it obvious} basis of a regular
design is not a Gr\"obner basis in general.
In fact, the right hand side of (\ref{eqn:basis-2^7-4}) is not a
Gr\"obner basis for any term order. In the arguments above we used
the elimination theory as a general method to obtain a basis of the
design ideal and obtained  a reduced Gr\"obner basis as a result. 
However, it is important {\it in itself} to obtain a
Gr\"obner basis, which we see in the next section.

\section{The confounding relation and the ideal membership problem}
\label{sec:confounding-ideal-membership}
In this section we see from the Gr\"obner bases theory that the
confounding relation can be
generalized from regular to non-regular designs and expressed
concisely. This is one of the merits to consider the design ideal
$I({\cal F})$. In fact, the problem of judging whether two factor effects
or interaction effects are confounded or not is equivalent to the ideal
membership problem, which  is solved by calculation of a Gr\"obner basis of
the design ideal. We give an overview of this fact. 
For details
see \cite{Pistone-Wynn-1996} or \cite{Galetto-Pistone-Rogantin-2003}.

First we give some necessary notation and definitions. 
${\rm LT}_{\tau}(f)$ denotes the leading term of the polynomial $f \in
K[x_1,\ldots,x_m]$ with respect to the term order $\tau$.
For an ideal $I \subset K[x_1,\ldots,x_m]$, we write 
the set of the leading terms of the elements in $I$ as 
${\rm LT}_{\tau}(I) = \{{\rm LT}_{\tau}(f)\ |\ f \in I\}$.
A monomial is called a standard monomial if it does not belong to 
${\rm LT}_{\tau}(I)$. 
{}From the definition of Gr\"obner
basis, the set of standard monomials
 is also characterized as the set of monomials which is
not divisible by any leading term of the element of the Gr\"obner basis with
respect to the term order $\tau$. 
We write the set of standard monomials 
$\{\Bx^{\Ba}\ |\ \Bx^{\Ba} \not\in {\rm LT}_{\tau}(I({\cal F}))\}$
as ${\rm Est}_{\tau}({\cal F})$. 
The following is a 
basic theorem (Proposition 1.1 of \cite{Sturmfels-1996}) in the theory of Gr\"obner bases.

\begin{theorem}
\label{th:basis}
$K[x_1,\ldots,x_m]/I({\cal F})$ is isomorphic as a
 $K$-vector space to ${\rm Span}({\rm Est}_{\tau}({\cal F}))$.
${\rm Est}_{\tau}({\cal F})$ is a basis of this vector space.
\end{theorem}

${\rm Est}_{\tau}({\cal F})$
represents one of the identifiable sets of main and interaction effects
under the design $\cal F$
and 
the number of the monomials in 
$\mbox{Est}_{\tau}({\cal F})$ is always the same as the run size $n$ for
any $\tau$.

\begin{example}
For the two reduced Gr\"obner bases in Example 
\ref{exam:2^7-4}, ${\rm Est}_{\tau}({\cal F})$ is written as follows.
\begin{itemize}
\item lexicographic:\ $\{1, x_5, x_6, x_7, x_5x_6, x_5x_7, x_6x_7, x_5x_6x_7\}$
\item graded reverse lexicographic:\ $\{1, x_1, x_2, x_3, x_4, x_5, x_6, x_7\}$
\end{itemize}
\end{example}

Now we consider the relation between the confounding relation and the
design ideal. We identify a monomial 
$\Bx^{\Ba}$ with a {\it main effect} if 
$\sum_{i = 1}^m a_i = 1$, and a {\it two-factor interaction effect} if 
$\sum_{i = 1}^m a_i= 2$ and so on. Then two main or interaction effects
are confounded in the design ${\cal F}$ if 
$\Bx^{\Ba_1}\Bx^{\Ba_2}$ is identically equal to $+1$ (or $-1$) for all the points
in $\Bx \in {\cal F}$. The confounded effects cannot be estimated
simultaneously. Therefore a design has to be chosen such that main
effects are confounded with  higher order interaction effects under the
hierarchical assumption. This is the concept of resolution.
The confounding relation is expressed in terms of the design ideal
as follows.
\begin{proposition}
\label{prop:ideal-membership}
Let $c \in \{-1,+1\}$. Then the following two conditions are equivalent.
\[
 {\rm (i)}\ \ \Bx^{\Ba_1}\Bx^{\Ba_2} = c\ \ \mbox{\rm for all}\ \ \Bx \in {\cal F}
\hspace*{10mm}{\rm (ii)}\ \ \Bx^{\Ba_1} - c\Bx^{\Ba_2} \in I({\cal F})
\]
\end{proposition}
In general, we have to calculate a  Gr\"obner basis to judge whether a
given polynomial belongs to a given ideal or not, i.e., to solve the ideal
membership problem.

\begin{example} 
Consider the design in Example \ref{exam:2^7-4} again.
Since the defining relation $x_3=-x_1x_2$ exists, the main effect of
 $x_3$ and two-factor interaction effect of 
 $x_1$ and $x_2$ are confounded. 
 For the reduced Gr\"obner  basis (\ref{eqn:gb-grevlex}),
 $x_1x_2 + x_3 \in I({\cal F})$ is obvious since the basis includes $x_1x_2+x_3$. 
 On the other hand for the  reduced Gr\"obner basis (\ref{eqn:gb-lex}), it is shown as follows
\[
 x_1x_2 + x_3 = (x_1+x_6x_7)x_2 - (x_2+x_5x_7)x_6x_7 + (x_7^2-1)x_5x_6 +
 (x_3+x_5x_6).
\]
\end{example}

For the case that the factor has $s\ (s > 2)$ levels, similar relation
as in Proposition \ref{prop:ideal-membership}
holds if we code the levels as the $s$th root of unity.
For example of three-level factors, the
levels are coded as $\{1,\omega,\omega^2\},\ \omega =\exp(2\pi i/3)$.
See \cite{Pistone-Rogantin-2008b} and \cite{Aoki-Takemura-2009b} for details.

\section{Indicator functions}
\label{sec:indicator-function}
In this section, we introduce an {\it indicator function}, which is
defined by \cite{Fontana-Pistone-Rogantin-2000}. 
The indicator function of a design ${\cal F} \subset {\cal D}$ is a
polynomial $f \in K[x_1,\ldots,x_m]$ satisfying
\[
 f(\Bx) = \left\{\begin{array}{cl}
1, & \mbox{if}\ \Bx \in {\cal F}\\
0, & \mbox{if}\ \Bx \in {\cal D} \setminus {\cal F}.
\end{array}
\right.
\]
The indicator function has a unique square-free representation under the
constraint $x_i^2 = 1, i = 1,\ldots,m$, and in a one-to-one
correspondence to the design ${\cal F}$.
Many important results in the field of computational algebraic
statistics are related to the indicator function. For example,
some classes of fractional factorial designs can be classified 
by the coefficients of their indicator functions. It is also shown that
some concepts of designed experiments such as confounding,
resolution, orthogonality
and estimability, are related
to the structure of the indicator function of a design.
Since the indicator function is  defined for any
design,  some classical notions for regular designs, such as confounding and resolution,
can be generalized to non-regular designs naturally by the notion of the
indicator function. See \cite{Fontana-Pistone-Rogantin-2000} or 
\cite{Ye-2003} for details.

In addition, since the indicator function is a polynomial, it can be
incorporated into the theory of computational algebraic statistics
naturally. For example, the design ideal
$I({\cal F})$ is simply written as
\[
 I({\cal F}) = \langle x_1^2-1,\ldots,x_m^2-1,f(\Bx)-1
\rangle ,
\]
where $f(\Bx)$ is the indicator function of ${\cal F}$. In other words,
the indicator function forms a set of defining equations by itself.

We list  some characteristics of the indicator function. The indicator
function of the full factorial design 
${\cal D}$ is $f(\Bx) \equiv 1$. The constant term of the indicator function
of a fractional factorial designs is equal to the fraction $n/2^m$. 
The indicator function of a regular fractional factorial
designs is simply written as a product of its defining relations
(see \cite{Fontana-Pistone-Rogantin-2000} or \cite{Ye-2003}).
For example, the indicator function of the 
$2^{7-4}$ design in Example \ref{exam:2^7-4} is written as
\begin{equation}
\label{eq:regular-indicator-function}
 f(\Bx) = \frac{1}{16}(1-x_1x_2x_3)(1-x_1x_4x_5)(1-x_2x_4x_6)(1+x_1x_2x_4x_7).
\end{equation}
The absolute values of the coefficients in the indicator function do not
exceed the constant term. In particular, the absolute values of all the
coefficients in the indicator function of regular designs coincide with 
the constant term.

Concerning non-regular designs,
one of the results on the coefficients of the
indicator function of a non-regular design
is related to the existence of 
a regular design including the non-regular design.

\begin{example}[The indicator function of the non-regular fractional
 factorial designs] 
Consider the following three fractional factorial designs.
\[
 \begin{array}{rrr}
{\cal F}_1\\
x_1 & x_2 & x_3\\ \hline
1 & 1 & 1\\
1 & -1 & -1\\
-1 & 1 & -1\\
-1 & -1 & 1
 \end{array}
\hspace*{10mm}
 \begin{array}{rrr}
{\cal F}_2\\
x_1 & x_2 & x_3\\ \hline
1 & 1 & 1\\
1 & -1 & -1\\
-1 & 1 & -1\\
\multicolumn{3}{c}{}
 \end{array}
\hspace*{10mm}
 \begin{array}{rrr}
{\cal F}_3\\
x_1 & x_2 & x_3\\ \hline
1 & 1 & 1\\
1 & 1 & -1\\
1 &-1 & 1\\
-1 & 1 & 1
 \end{array}
\]
${\cal F}_1$ is a $2^{3-1}$ regular design defined by 
$x_1x_2x_3 = 1$, and ${\cal F}_2$ is a non-regular design which is a
 proper subset of ${\cal F}_1$. ${\cal F}_3$ is also a non-regular
 design, but there does not exist a regular design which includes ${\cal
 F}_3$ as a proper subset. The indicator functions of these three
 designs are given as follows.
\begin{align*}
{\cal F}_1:\ & f(\Bx) = \displaystyle\frac{1}{2} + \frac{1}{2}x_1x_2x_3\\
{\cal F}_2:\ & f(\Bx) = \displaystyle\frac{3}{8} +
 \frac{1}{8}(x_1+x_2+x_3-x_1x_2+x_1x_3+x_2x_3)+\frac{3}{8}x_1x_2x_3\\
{\cal F}_3:\ & f(\Bx) = \displaystyle\frac{1}{2} +
 \frac{1}{4}(x_1+x_2+x_3 - x_1x_2x_3)
\end{align*}
An important observation is that the terms whose coefficients are equal to to the
 constant term in the indicator function of ${\cal F}_2$ (i.e.,
 $x_1x_2x_3$) coincide with the terms
 of the indicator
 function of ${\cal F}_1$, and there are no such terms in the indicator
 function of ${\cal F}_3$.  These characteristics of indicator functions
 hold in general.
\end{example}

The absolute value  of a coefficient in the indicator function
represents a complete confounding relation if it is equal to the constant term,
and a partial confounding relation if it is smaller than the
constant term. Considering this point, we gave in \cite{Aoki-Takemura-2009a}
a new class of two-level non-regular fractional factorial designs, called an {\it
affinely full-dimensional factorial design}, which has a desirable
property for the identifiability of parameters.
We present it briefly in Section \ref{sec:opt-design-selection}.

\section{Indicator function for adding factors}
\label{sec:adding-factor}
In this section, as a simple application of the indicator function, we
consider the design ideal for adding factors. 
The additional factors may be real controllable factors, whose
levels are determined by some defining relations.  For the purpose of
Markov bases in Section \ref{sec:markov-basis} the additional factors
are formal and correspond to interaction effects included in a
null hypothesis. 

Let ${\cal F}_1$ be a fractional factorial design of the factors
$x_1,\ldots,x_m$. Consider adding 
factors $y_1,\ldots,y_k$ to ${\cal F}_1$.
We suppose the levels of the additional factors are determined by the
defining relations among $x_1,\ldots,x_m$ as 
\[
 y_1 = e_1\Bx^{\Bb_1},\ldots,y_k = e_k\Bx^{\Bb_k},
\]
where $e_1,\ldots,e_k \in \{-1,1\}$. Write this new design of
$x_1,\ldots,x_m,y_1,\ldots,y_k$ as ${\cal F}_2$. The run sizes of ${\cal
F}_1$ and ${\cal F}_2$ are the same.

Let $f_1$ and $f_2$ be the indicator functions of ${\cal F}_1$ and ${\cal
F}_2$, respectively. Then we have
\begin{equation}
\label{eq:adding-factor}
 f_2(x_1,\ldots,x_m,y_1,\ldots,y_k) =
      \frac{1}{2^k}(1+e_1y_1\Bx^{\Bb_1})\cdots(1 +
      e_ky_k\Bx^{\Bb_k})f_1(x_1,\ldots,x_m).
\end{equation}
In fact, for $(x_1,\ldots,x_m,y_1,\ldots,y_k) \in {\cal F}_2$, 
$(x_1,\ldots,x_m) \in {\cal F}_1$ and 
\[
 e_1y_1\Bx^{\Bb_1} = \cdots = e_ky_k\Bx^{\Bb_k} = 1
\]
hold, which yields $f_2 = 1$. Conversely, if 
$(x_1,\ldots,x_m,y_1,\ldots,y_k) \not\in {\cal F}_2$, then 
$(x_1,\ldots,x_m) \not\in {\cal F}_1$ or some of
$e_1y_1\Bx^{\Bb_1},\ldots,e_ky_k\Bx^{\Bb_k}$ has to be $-1$, which yields 
$f_2 = 0$. Note that 
(\ref{eq:adding-factor}) generalizes the indicator function 
of regular fractional factorial designs 
(\ref{eq:regular-indicator-function}), by 
taking $f_1\equiv 1$, i.e.\ by assuming the full factorial design for 
$x_1, \dots, x_m$. 

From the above result, we have an expression of $I({\cal F}_2)$
\[
 I({\cal F}_2) =
      \langle x_1^2-1,\ldots,x_m^2-1,y_1^2-1,\ldots,y_k^2-1,f_1-1,f_2-1
\rangle .
\]
If we fix the term order $\tau$ on $x_1,\ldots,x_m$ and  
 $\sigma$ on $x_1,\ldots,x_m,y_1,\ldots,y_k$, 
$\mbox{Est}_{\tau}({\cal F}_1)$ and $\mbox{Est}_{\sigma}({\cal F}_2)$
are defined. 
$\mbox{Est}_{\tau}({\cal F}_1)$ and $\mbox{Est}_{\sigma}({\cal F}_2)$
contain the same number of monomials since the run sizes of 
${\cal F}_1$ and ${\cal F}_2$ are the same. In particular, if we
use a term order $\sigma$ such that
$\{y_1,\ldots,y_k\} \succ_\sigma \{x_1,\ldots,x_m\}$, then
$\mbox{Est}_{\tau}({\cal F}_1) = \mbox{Est}_{\sigma}({\cal F}_2)$ holds.

\section{Consideration of selection of optimal designs}
\label{sec:opt-design-selection}
This section is based on two works by the authors 
related to optimal selection of non-regular designs. 

In \cite{Aoki-Takemura-2009a} we defined a new class of two-level
non-regular fractional factorial designs as follows.
\begin{definition}[Definition 2.1 of \cite{Aoki-Takemura-2009a}]
A non-regular fractional factorial design ${\cal F}$ is called an affinely
 full-dimensional factorial design if there is no regular fractional
 factorial design ${\cal F}'$ satisfying ${\cal F}
 \subsetneq {\cal F}'$. Conversely,
a non-regular fractional factorial design ${\cal F}$ is called a subset
 fractional factorial design if there is some regular fractional
 factorial design ${\cal F}'$ satisfying ${\cal F} \subsetneq {\cal F}'$.
\end{definition}
At a glance, the merit of this definition might not be clear. One of the
properties of the affinely full-dimensional factorial design is the
simultaneous identifiability of the parameters. In fact, if ${\cal F}$
is an affinely full-dimensional factorial design, then the parameters
of the main effect model are simultaneously identifiable and vice
versa. See Lemma 2.1 and Lemma 2.2 of \cite{Aoki-Takemura-2009a}.
As another interesting property of the affinely full-dimensional
factorial design, we have the following conjecture 
(Conjecture 3.1 of \cite{Aoki-Takemura-2009a})
on the $D$-optimality of affinely full-dimensional 
designs.

\begin{conjecture}
Consider the main effect model for the observations obtained in a
fractional factorial design of $m$ factors. Then 
$D$-optimal design is affinely full-dimensional factorial if and only if  
$m = 5,6,7$ (mod $8$).
\end{conjecture}
Though this conjecture is not proved in general, it is shown to be true
when some known bounds for the maximal determinant problem
(\cite{Barba-1933}, \cite{Ehlich-1964} and
\cite{Wojtas-1964}) are attained. See \cite{Aoki-Takemura-2009a} for details.

In \cite{Aoki-2010}, we considered more realistic situation that 
the model is unknown. In this case, we cannot rely on model-based criterion
such as $D$-optimality and have to 
evaluate the model-robustness. In \cite{Aoki-2010}, 
we considered the situation
where (i) all the main effects are of primary interest and their
estimates are required, (ii) an experimenter assumes that there are certain number
of active two-factor interaction effects and certain number of 
active three-factor interaction effects, 
but it is unknown which of two- and three-factor interactions are active,
and (iii) all the four-factor and higher-order
interactions are negligible. This is a natural extension of the setting
considered in \cite{Cheng-Deng-Tang-2002}.
In \cite{Aoki-Takemura-2009a}, 
we presented some optimality criteria to evaluate model-robustness of
 non-regular two-level fractional factorial designs.
Our approach was 
 based on minimizing the sum
 of squares of all the off-diagonal elements in the information
 matrix, and considering expectation under appropriate distribution functions
 for unknown contamination of the interaction effects. 
We also compared our criterion to a generalized minimum aberration
criterion by \cite{Deng-Tang-1999}
and affinely full dimensionality. 

\section{Markov bases and conditional tests by Markov chain Monte Carlo
 method}
\label{sec:markov-basis}
In this section we introduce another topic of application of Gr\"obner basis
theory to the designed experiments, called {\it Markov bases for designed
experiments}. The  notion of Markov bases was  introduced in 
\cite{Diaconis-Sturmfels-1998}. They established a procedure for sampling
from discrete conditional distributions by constructing a connected
Markov chain on a given sample space. Since
then many works have been published on the topic of Markov bases by both
algebraists and statisticians. 
This constitutes another main branch of the field of computational
algebraic statistics. See \cite{Aoki-Takemura-2009-trans} for
the history of this topic. 
It is of interest to investigate statistical problems which are related
to both designed experiments and Markov bases. 
In \cite{Aoki-Takemura-2010} and
\cite{Aoki-Takemura-2009b} we considered applying Markov bases
for discrete observations from designed experiments. 
In this section we review the results of these works.

Suppose we have nonnegative integer  observations for each run of a
regular fractional design.
For simplicity, we also suppose that the observations are
counts of some events and only one observation is obtained for each
run. In this case it is natural to consider the Poisson model,
in the framework of generalized linear models
(\cite{McCullagh-Nelder-1989}). 
Write the observations as $\By = (y_1,\ldots,y_n)'$, where $n$ is
the run size and $'$ denotes the transpose. 
The observations are realizations from $n$ Poisson random
variables $Y_1,\ldots,Y_n$, which are mutually independently distributed
with the mean parameter $\mu_i = E(Y_i), i = 1,\ldots,n$. We express the
mean parameter $\mu_i$ as
\begin{equation}
 \log \mu_i = \beta_0 + \beta_1 x_{i1} + \cdots + \beta_{\nu-1}x_{i\nu-1},
\label{eqn:log-linear-null}
\end{equation}
where $\Bbeta = (\beta_0,\beta_1,\ldots,\beta_{\nu-1})'$ is the
$\nu$-dimensional parameter and $x_{i1},\ldots,x_{i\nu-1}$ are the
$\nu-1$ covariates. We write the covariate matrix $A$ as
\begin{equation}
 A = \left(\begin{array}{cccc}
1 & x_{11} & \cdots & x_{1\nu-1}\\
\vdots & \vdots & \cdots & \vdots\\
1 & x_{n1} & \cdots & x_{n\nu-1}\\
\end{array}
\right). 
\label{eqn:covariate-matrix}
\end{equation}
Note that the expression (\ref{eqn:log-linear-null}) can be treated as
the {\it null model} $\mbox{H}_0$.  Since the 
saturated model has $n$-dimensional parameter, various goodness-of-fit
tests with the saturated model as the alternative $\mbox{H}_1$  can be written as
\[
 \begin{array}{l}
\mbox{H}_0:\ (\beta_{\nu},\ldots,\beta_n) = (0,\ldots,0)\\
\mbox{H}_1:\ (\beta_{\nu},\ldots,\beta_n) \neq (0,\ldots,0)
\end{array}
\]
by introducing additional parameter $(\beta_{\nu},\ldots,\beta_n)$. 
Various other hypotheses can be written in a similar way.
Under the null model  (\ref{eqn:log-linear-null})
the  sufficient statistic for the parameter $\Bbeta$
is given by 
$A'\By = (\sum_{i=1}^n y_i, \sum_{i = 1}^n x_{i1}y_i, \ldots, \allowbreak\sum_{i = 1}^n x_{i\nu-1}y_i)'$.

Once we specify the null model and a test statistic, our purpose is to
calculate the $p$ value. In this stage, Markov chain Monte Carlo
procedure is a valuable tool, especially when the fitting of the traditional
large-sample approximation is poor and the exact calculation of the $p$
value is infeasible. To perform the Markov chain Monte Carlo procedure,
the key notion is a Markov basis over the conditional sample space
given the values of the sufficient statistic
\begin{equation}
\{\By\ |\ A'\By = A'\By^o,\ y_i\ \mbox{is a non-negative integer},\ i =
1,\ldots,n\},
\label{eqn:fiber}
\end{equation}
where $\By^o$ is the observed  vector. Once a Markov basis is
calculated, we can construct a connected, aperiodic and reversible
Markov chain over the space in (\ref{eqn:fiber}), which can be modified
so that the stationary distribution is the conditional distribution
under the null model by the Metropolis-Hastings procedure. See
\cite{Diaconis-Sturmfels-1998} and \cite{Hastings-1970} for details.

In the arguments above, an important step is to construct a covariate
matrix $A$ in (\ref{eqn:covariate-matrix}). In this step, we have to
construct $A$ so that all the parameters in (\ref{eqn:log-linear-null})
are simultaneously estimable. This problem corresponds to the ideal
membership problem in Section \ref{sec:confounding-ideal-membership}.

We illustrate the above setup with two examples.

We first consider 
a $2^{7-3}$ fractional factorial design.
Suppose we have observations $\By = (y_1,\ldots,y_{16})'$ for each run of
 the fractional factorial design with the defining relation
\begin{equation}
 x_1x_2x_4x_5 = x_1x_3x_4x_6 = x_2x_3x_4x_7 = 1.
\label{eqn:example-def-relation}
\end{equation}
The design and the observation is written as follows.
\[
{\footnotesize
\begin{array}{rrrrrrrrc}\hline
& \multicolumn{7}{c}{\mbox{Factor}} & \By\\
\mbox{Run} & x_1 & x_2 & x_3 & x_4 & x_5 & x_6 & x_7 &  \\ \hline
 1 & 1 & 1 & 1 & 1 & 1 & 1 & 1 & y_1\\
 2 & 1 & 1 & 1 & -1 & -1 & -1 & -1 & y_2\\
 3 & 1 & 1 & -1 & 1 & 1 & -1 & -1 & y_3\\
 4 & 1 & 1 & -1 & -1 & -1 & 1 & 1 & y_4\\
 5 & 1 & -1 & 1 & 1 & -1 & 1 & -1 & y_5\\
 6 & 1 & -1 & 1 & -1 & 1 & -1 & 1 & y_6\\
 7 & 1 & -1 & -1 & 1 & -1 & -1 & 1 & y_7\\
 8 & 1 & -1 & -1 & -1 & 1 & 1 & -1 &  y_8\\
 9 & -1 & 1 & 1 & 1 & -1 & -1 & 1 & y_9\\
10 & -1 & 1 & 1 & -1 & 1 & 1 & -1 & y_{10}\\
11 & -1 & 1 & -1 & 1 & -1 & 1 & -1 & y_{11}\\
12 & -1 & 1 & -1 & -1 & 1 & -1 & 1 & y_{12}\\
13 & -1 & -1 & 1 & 1 & 1 & -1 & -1 & y_{13}\\
14 & -1 & -1 & 1 & -1 & -1 & 1 & 1 & y_{14}\\
15 & -1 & -1 & -1 & 1 & 1 & 1 & 1 & y_{15}\\
16 & -1 & -1 & -1 & -1 & -1 & -1 & -1 & y_{16}\\
\end{array}
}
\]
In this case, there are several models to be considered. If our interest is
only on the main effects for seven factors, we may define
\[
\Bbeta = (\beta_0, \beta_1, \ldots, \beta_7)'
\] 
and
\begin{equation}
\label{eq:covariate-transpose}
A= 
\left(\begin{array}{rrrrrrrrrrrrrrrr}
 1 & 1 & 1 & 1 & 1 & 1 & 1 & 1 & 1 & 1 & 1 & 1 & 1 & 1 & 1 & 1\\
 1 & 1 & 1 & 1 & 1 & 1 & 1 & 1 &-1 &-1 &-1 &-1 &-1 &-1 &-1 &-1\\
 1 & 1 & 1 & 1 &-1 &-1 &-1 &-1 & 1 & 1 & 1 & 1 &-1 &-1 &-1 &-1\\
 1 & 1 &-1 &-1 & 1 & 1 &-1 &-1 & 1 & 1 &-1 &-1 & 1 & 1 &-1 &-1\\
\vdots & \vdots & \vdots & \vdots & \vdots & \vdots & \vdots & \vdots & 
\vdots & \vdots & \vdots & \vdots & \vdots & \vdots & \vdots & \vdots\\
 1 &-1 &-1 & 1 & 1 &-1 &-1 & 1 &-1 & 1 & 1 &-1 &-1 & 1 & 1 &-1\\
 1 &-1 &-1 & 1 &-1 & 1 & 1 &-1 & 1 &-1 &-1 & 1 &-1 & 1 & 1 &-1
\end{array}\right)'
\end{equation}
i.e., the covariate matrix $A$ is constructed as the design matrix and the 
column vector $(1,\ldots,1)'$. In this case, 
the parameter $\beta_j$ is interpreted as the parameter contrast for the
main effect of the factor $x_j$ for $j = 1,\ldots,7$.

We can also consider models containing interaction effects. For example,
if we want to estimate the two-factor interaction effect among $x_1$ 
and $x_2$  along with all
the main effects, we may add the column
\begin{equation}
(1,1,1,1,-1,-1,-1,-1,-1,-1,-1,-1,1,1,1,1)'
\label{eqn:add-inter-12}
\end{equation}
to the above $A$. 
Note that this corresponds to adding a factor in Section \ref{sec:adding-factor}.
In this case, $\Bbeta$ is $9$-dimension and the 
element corresponding to the additional row (\ref{eqn:add-inter-12})
is interpreted as the parameter contrast for the two factor interaction
effect among $x_1$ and $x_2$. 

If we want to estimate another interaction, we have to consider the 
confounding relations. 
For example, two two-factor interaction effects among 
$x_1 \times x_2$ and $x_4 \times x_5$ cannot be estimated simultaneously
since they are confounded. This confounding relation is 
shown in (\ref{eqn:example-def-relation}). In the terms of algebra, 
this confounding relation is shown as the ideal membership, 
\[
x_1x_2 - x_4x_5 \in \langle 
x_1^2-1,\ldots,x_7^2-1,x_1x_2x_4x_5-1,x_1x_3x_4x_6-1,x_2x_3x_4x_7-1
\rangle ,
\]
as discussed  in Section \ref{sec:confounding-ideal-membership}. 


As our second example,
we consider the case of three-level factors.  We also indicate
how complex coding simplifies the specification of 
the conditional sample space (\ref{eqn:fiber}).
Consider the following $3^{3-1}$ fractional factorial design, where
the levels are coded as $\{0,1,2\}$.
\[
\begin{array}{rrrrc}\hline
& \multicolumn{3}{c}{\mbox{Factor}} & \By\\
\mbox{Run} & x_1 & x_2 & x_3 &  \\ \hline
 1 & 0 & 0 & 0 & y_1\\
 2 & 0 & 1 & 2 & y_2\\
 3 & 0 & 2 & 1 & y_3\\
 4 & 1 & 0 & 2 & y_4\\
 5 & 1 & 1 & 1 & y_5\\
 6 & 1 & 2 & 0 & y_6\\
 7 & 2 & 0 & 1 & y_7\\
 8 & 2 & 1 & 0 & y_8\\
 9 & 2 & 2 & 2 & y_9\\
\end{array}
\]
In this case, each main and interaction factor has more than one degree of
freedom and has to be parameterized by more than one parameters. For example,
the main effect of $x_1$ can be expressed $(\alpha_1, \alpha_2, \alpha_3)$
with one constraint. One of the simplest constraints is $\alpha_3=0$, i.e.,
to treat $x_3$ as the baseline, which is expressed as the two columns
\[\left(
\begin{array}{ccccccccc}
1 & 1 & 1 & 0 & 0 & 0 & 0 & 0 & 0\\
0 & 0 & 0 & 1 & 1 & 1 & 0 & 0 & 0\\
\end{array}
\right)'\]
in the covariate matrix $A$. 
We can also consider a symmetric constraint
$\alpha_1+\alpha_2+\alpha_3=0$. In this case, we may include the two columns
\[\left(
\begin{array}{ccccccccc}
2 & 2 & 2 & 0 & 0 & 0 & -1 & -1 & -1\\
0 & 0 & 0 & 2 & 2 & 2 & -1 & -1 & -1\\
\end{array}
\right)'\]
to $A$. Note that the conditional sample space (\ref{eqn:fiber})
is invariant for the constraints since the total $\sum y_i$ is fixed.
As for the interaction effects, see \cite{Aoki-Takemura-2009b}.

Another equivalent expression is given directly from the complex coding
\[
\begin{array}{ccccc}\hline
& \multicolumn{3}{c}{\mbox{Factor}} & \By\\
\mbox{Run} & x_1 & x_2 & x_3 &  \\ \hline
 1 & 1 & 1 & 1 & y_1\\
 2 & 1 & \omega & \omega^2 & y_2\\
 3 & 1 & \omega^2 & \omega & y_3\\
 4 & \omega & 1 & \omega^2 & y_4\\
 5 & \omega & \omega & \omega & y_5\\
 6 & \omega & \omega^2 & 1 & y_6\\
 7 & \omega^2 & 1 & \omega & y_7\\
 8 & \omega^2 & \omega & 1 & y_8\\
 9 & \omega^2 & \omega^2 & \omega^2 & y_9\\
\end{array}
\]
where 
$\omega=\exp(2\pi i/3)$.
If we allow $A$ to be a complex matrix and consider the real and the
complex parts, 
the conditional sample space (\ref{eqn:fiber}) defined from 
\[
A = \left(\begin{array}{cccc}
 1 & 1 & 1 & 1 \\
 1 & 1 & \omega & \omega^2\\
 1 & 1 & \omega^2 & \omega\\
 1 & \omega & 1 & \omega^2\\
 1 & \omega & \omega & \omega\\
 1 & \omega & \omega^2 & 1\\
 1 & \omega^2 & 1 & \omega\\
 1 & \omega^2 & \omega & 1\\
 1 & \omega^2 & \omega^2 & \omega^2
\end{array}
\right)
\]
is also the same. 
This follows from the following  basic fact of the theory of cyclotomic polynomials
(e.g.\ Section 6.3 of \cite{lang-algebra-3rd}):
for $q_1, q_2, q_3\in \mathbb{Q}$
\[
q_1 + q_2 \omega + q_3 \omega^2 = 0 \ \Leftrightarrow \ q_1 = q_2 = q_3.
\]
The same result holds for $s$-level factors, where $s$ is a prime number.

\section{Some discussions}
\label{sec:discussions}

In this paper, we review several topics related to  designed
experiments and computational algebraic statistics. As we have stated,
there are two works as the beginning of  computational algebraic
statistics, i.e., the work by Pistone and Wynn
(\cite{Pistone-Wynn-1996}) and by Diaconis and Sturmfels
(\cite{Diaconis-Sturmfels-1998}). 
It is important to study whether
a closer connection can be established
between these two branches, i.e., 
the design ideal and the Markov basis (toric
ideal). Our works \cite{Aoki-Takemura-2009b} and
\cite{Aoki-Takemura-2010} are motivated by this goal,
although we do not yet have a result of some general nature.

The following argument suggests that there 
should be some general results relating these two branches.
Recall the covariate matrix $A$ in (\ref{eq:covariate-transpose}).
For the main effect model, except for the first element $1$, each row of 
$A$ is just a point in the design and therefore $A$ itself 
can be considered as a design.  In the theory of toric ideals, the
set of rows of $A$ is often called a configuration defining the toric ideal.
Therefore for the main effect model, the design
simultaneously defines the design ideal and  the toric ideal.
Note that this relation also holds for three-level 
factors (or more generally for prime number of levels),
if the levels are coded by complex numbers as indicated
at the end of Section \ref{sec:markov-basis}.

If we include some interaction effects in the null model for two-level case, this 
corresponds to adding factors as in Section \ref{sec:adding-factor}.
Therefore again there is a very simple relation between the
design (without added factors) and the configuration for the toric ideal 
(with added factors).   This argument suggests that some
algebraic properties of the design ideal should be reflected
in algebraic properties of the toric ideal.  This is an important
topic for further research.

\bibliographystyle{plain}

\begin{thebibliography}{99}
\bibitem{Adams-Loustaunau-1994}
W. W. Adams and P. Loustaunau (1994).
{\it An Introduction to Gr\"obner Bases}. Graduate Studies in
	Mathematics. American Mathematical Society, Providence, RI.
\bibitem{Aoki-2010}
S. Aoki (2010).
Some optimal criteria of model-robustness for two-level non-regular
	fractional factorial designs. {\it Annals of the Institute of
	Statistical Mathematics}, to appear.
\bibitem{Aoki-Takemura-2009a}
S. Aoki and A. Takemura (2009a).
Some characterizations of affinely full-dimensional factorial
	designs. {\it Journal of Statistical Planning and Inference},
	{\bf 139}, 3525--3532.
\bibitem{Aoki-Takemura-2009b}
S. Aoki and A. Takemura (2009b).
Markov basis for design of experiments with three-level factors. in {\it
	Algebraic and Geometric Methods in Statistics} (dedicated to
	Professor Giovanni Pistone on the occasion of his sixty-fifth
	birthday), edited by P. Gibilisco, E. Riccomagno,
	M. P. Rogantin and H. P. Wynn, Cambridge University Press,
	225--238.
\bibitem{Aoki-Takemura-2009-trans}
S. Aoki and A. Takemura (2009c).
Statistics and Gr\"obner bases --- The origin and development of
	computational algebraic statistics, in {\it Selected Papers on
	Probability and Statistics}. American Mathematical Society
	Translations. Series 2, Volume 227, 125--145.
\bibitem{Aoki-Takemura-2010}
S. Aoki and A. Takemura (2010).
Markov chain Monte Carlo tests for designed experiments. {\it Journal of
	Statistical Planning and Inference}, {\bf 140}, 817--830.
\bibitem{Barba-1933} G. Barba (1933). Intorno al teorema di Hadamard sui
        determinanti a valore massimo, {\it Giorn. Mat. Battaglini},
        {\bf 71}, 70--86.
\bibitem{Box-Hunter-1961}
G. E. P. Box and J. S. Hunter (1961).
The $2^{k-p}$ fractional factorial design. {\it Technometrics}, {\bf 3}, 
311--351, 449--458.
\bibitem{Box-Hunter-Hunter-1978}
G. E. P. Box, W. C. Hunter and J. S. Hunter (1978).
{\it Statistics for Experiments}, Wiley, New York.
\bibitem{Cheng-Deng-Tang-2002}
Cheng, C. S., Deng, L. Y. and Tang, B. (2002).
Generalized minimum aberration and design efficiency for nonregular
        fractional factorial designs. {\it Statistica Sinica}, {\bf 12},
        991--1000.
\bibitem{Cox-Little-O'Shea-1997}
D. Cox, J. Little and D. O'Shea (2007).
{\it Ideal, Varieties, and Algorithms}, 3rd edition. Springer, New York.
\bibitem{Deng-Tang-1999}
L. Y. Deng and B. Tang (1999).
Generalized resolution and minimum aberration criteria for
        Plackett-Burman and other nonregular factorial designs.
{\it Statistica Sinica}, {\bf 9}, 1071--1082.


\bibitem{Diaconis-Sturmfels-1998}
P. Diaconis and B. Sturmfels (1998).
Algebraic algorithms for sampling from conditional distributions. {\it
	Annals of Statistics}, {\bf 26}, 363--397.
\bibitem{Ehlich-1964} H. Ehlich (1964). Determinantenabsch\"atzungen
        f\"ur bin\"are Matrizen, {\it Math. Z.}, {\bf 83}, 123--132.
\bibitem{Fontana-Pistone-Rogantin-2000}
R. Fontana, G. Pistone and M. P. Rogantin (2000). Classification of
	two-level factorial fractions. {\it Journal of Statistical
	Planning and Inference}, {\bf 87}, 149--172.
\bibitem{Galetto-Pistone-Rogantin-2003}
F. Galetto, G. Pistone and M. P. Rogantin (2003). Confounding revisited
	with commutative computational algebra. {\it Journal of
	Statistical Planning and Inference}, {\bf 117}, 345--363.
\bibitem{Hastings-1970}
W. K. Hastings (1970).
Monte Carlo sampling methods using Markov chains and their
	applications. {\it Biometrika}, {\bf 57}, 97--109.
\bibitem{lang-algebra-3rd}
S. Lang (2002). {\it Algebra}. 3rd ed. 
 Graduate Texts in Mathematics, 211. Springer, New York.
\bibitem{McCullagh-Nelder-1989}
P. McCullagh and J. A. Nelder (1989).
{\it Generalized linear models}. 2nd ed. Chapman \& Hall,  London.
\bibitem{Mukerjee-Wu-2006}
R. Mukerjee and C. F. J. Wu (2006).
{\it A Modern Theory of Factorial Designs}. Springer Series in Statistics.
\bibitem{Pistone-Riccomagno-Wynn-2001}
G. Pistone, E. Riccomagno and H. P. Wynn (2001).
{\it Algebraic Statistics: Computational Commutative Algebra in Statistics}. 
Chapman \& Hall, London.
\bibitem{Pistone-Rogantin-2008a}
G. Pistone and M. P. Rogantin (2008a).
Algebraic statistics of level codings for fractional factorial designs.
  {\it Journal of Statistical Planning and Inference}, 
  {\bf 138}, 234--244.
\bibitem{Pistone-Rogantin-2008b}
G. Pistone and M. P. Rogantin (2008b).
Indicator function and complex coding for mixed fractional factorial
	designs. {\it Journal of Statistical Planning and Inference},
	{\bf 138}, 787--802.
\bibitem{Pistone-Wynn-1996}
G. Pistone and H. P. Wynn (1996).
Generalised confounding with Gr\"obner bases. {\it Biometrika}, {\bf
	83}, 653--666.
\bibitem{Sturmfels-1996}
B. Sturmfels (1996). {\it Gr\"obner Bases and Convex
	Polytopes}. American Mathematical Society, Providence, RI.
\bibitem{Wojtas-1964} 
W. Wojtas (1964). On Hadamard's inequality for the
        determinants of order non-divisible by $4$, {\it Colloq. Math.},
        {\bf 12}, 73--83.
\bibitem{Ye-2003}
K. Q. Ye (2003). Indicator function and its application in two-level
	factorial designs. {\it Annals of Statistics}, {\bf 31},
	984--994.

\end{thebibliography}

\end{document}